\documentclass[fleqn,twoside]{article}
\usepackage{espcrc2}

\usepackage{graphicx}
\usepackage{epsfig}
\usepackage[figuresright]{rotating}

\def\etal{{\it et al.}}

\def\tm{\tau^-}
\def\tp{\tau^+}
\def\nut{\nu_{\tau}}

\def\gt{g_{\tau}}
\def\gm{g_{\mu}}
\def\ge{g_{\mathrm{e}}}
\def\gl{g_{l}}

\def\taue  {\tm \rightarrow \mbox{e}^- \bar{\nu}_{e}\nut }
\def\taum  {\tm \rightarrow \mu^- \bar{\nu}_{\mu} \nut }
\def\taul  {\tm \rightarrow l^- \bar{\nu}_{l} \nut }

\def\mue  {\mu^- \rightarrow \mbox{e}^- \bar{\nu}_e \nu_{\mu} }

\def\tauhpmay{\tau^- \rightarrow {\mbox {\rm h}}^- \geq 0\pi^0 \nu_{\tau}} 
\def\tauhhh{\tau^- \rightarrow \mathrm{h}^-\mathrm{h}^-\mathrm{h}^+ \geq 0\pi^0 \nu_{\tau}}

\def\ee{\mbox{e}^+\mbox{e}^-}
\def\mm{\mu^+\mu^-}
\def\tt{\tau^+\tau^-}
\def\qq{\mbox{q}\overline{\mbox{q}}}
\def\ggee{(\ee)\; \ee}
\def\ggmm{(\ee)\; \mm}

\def\reeee{\ee \rightarrow \ee}
\def\reemm{\ee \rightarrow \mm}
\def\reeqq{\ee \rightarrow \qq}

\def\reeeeee{\ee \rightarrow \ggee}
\def\reeeemm{\ee \rightarrow \ggmm}

\def\dedxt{\mathrm{d}E/\mathrm{d}x}

\def\ntks{N_{\mathrm{tks}}}
\def\pa{p_{\mathrm{1}}}
\def\pb{p_{\mathrm{2}}}
\def\pc{p_{\mathrm{3}}}
\def\popp{p_{\mathrm{1-opp}}}

\def\nmu{N_{\mathrm{muon}}}

\def\chisq{\chi^2}

\def\acol{\theta_{\mathrm{acol}}}
\def\mumatch{\mu_{\mathrm{match}}}
\def\ejet{E_{\mathrm{jet}}}

\def\btaum{B($\taum$)}

\newcommand{\AmS}{{\protect\the\textfont2
  A\kern-.1667em\lower.5ex\hbox{M}\kern-.125emS}}

\title{A Measurement of the $\taum$ Branching Ratio}

\author{L. Kormos\address{Department of Physics and Astronomy, 
        University of Victoria, \\ 
        P.O. Box 3055, STN CSC, Victoria BC V8W 3P6, Canada}}
       
\begin{document}

\begin{abstract}
The $\taum$ branching ratio has been measured using data collected
from 1990 to 1995 by the OPAL detector at the LEP collider.  The
resulting value of 
$\mbox{\btaum} = 0.1734 \pm 0.0009(stat) \pm 0.0005(syst)$
 has been used in conjunction with other OPAL
measurements to test lepton universality, yielding the coupling
constant ratios $\gm/\ge = 1.0005 \pm 0.0043$ and $\gt/\ge = 1.0031
\pm 0.0047$, in good agreement with the Standard Model prediction of
unity, and also to determine a
value for the Michel parameter $\eta = 0.004 \pm 0.036$.  This is
subsequently used to find a model-dependent limit of the mass for the
charged Higgs boson, $m_{\mathrm{H}^{\pm}} > 1.30 \tan\beta$,  in the Minimal
Supersymmetric Standard Model framework.
\vspace{1pc}
\end{abstract}

\maketitle

\section{Introduction}
Precise measurements of the leptonic decays of $\tau$ leptons provide
a means of stringently testing various aspects of the Standard Model.  OPAL previously has studied the leptonic $\tau$ decay modes by measuring the branching ratios  \cite{steve,opalt2m}, the Michel parameters \cite{rainer}, and radiative decays \cite{radt2m}.  This work presents a new OPAL measurement of the
$\taum$
branching ratio, using data taken from 1990 to 1995 at energies near the Z$^0$
peak, corresponding to an integrated luminosity of approximately 170
pb$^{-1}$.  A pure sample of $\tt$ pairs is selected from the LEP1 data
set as described in Section \ref{tausel}, and then the fraction of $\tau$
jets in which the $\tau$ has decayed to a muon is determined, using
the selection described in Section \ref{taumsel}.  This fraction is then
corrected for backgrounds, efficiency, and bias, as described in
Section \ref{br}.  The  selection of $\taum$ candidates relies on only a few variables, each of which provides a highly
effective means of separating background events from signal events
while minimising 
systematic uncertainty.  This new measurement supersedes the previous
OPAL measurement of B$(\taum) = 0.1736 \pm 0.0027$ which was obtained
using data collected in 1991 and 1992, corresponding to an integrated
luminosity of approximately 39 pb$^{-1}$ \cite{opalt2m}.  

        OPAL \cite{opal} is a general purpose detector covering almost
        the
        full solid angle with approximate cylindrical symmetry about
        the $\ee$ beam axis.  In the OPAL coordinate system, the $e^-$ beam
        direction defines the $+z$ axis, and the $+x$ axis points from the detector towards the centre of the LEP
        ring.  The polar angle $\theta$ is
        measured from the $+z$ axis, and the azimuthal angle $\phi$ is
        measured from the $+x$ axis. 

Selection efficiencies and kinematic variable distributions were
modelled using Monte Carlo simulated $\tp\tm$ event samples generated
with the KORALZ 4.02 package \cite{koralz} and the TAUOLA 2.0 library \cite{tauola}.  These events
were then passed through a full simulation of the OPAL detector \cite{opalsim}.
Background contributions from non-$\tau$ sources were evaluated using
Monte Carlo samples based on the following generators:  multihadronic
events $(\reeqq)$ were simulated using JETSET 7.3 and JETSET 7.4 \cite{jetset},
$\reemm$ events using KORALZ \cite{koralz}, Bhabha events using
BHWIDE \cite{bhwide}, and two-photon events using VERMASEREN \cite{vermaseren}.
\section{\label{tausel} The $\tp\tm$ selection}
        At LEP1, electrons and positrons were made to collide at
        centre-of-mass energies close to the Z$^0$ peak, producing
        Z$^0$ bosons at rest which 
        subsequently decayed into back-to-back pairs of leptons
        or quarks, from which the $\tp\tm$ pairs were selected
        for this analysis.  These highly relativistic  $\tau$
        particles  decay in
        flight close to the interaction point, resulting in two highly-collimated, back-to-back jets
        in the tracking chamber. 

The $\tp\tm$ selection requires that an event have two
        jets as defined by the cone algorithm in reference
        \cite{jetalgo}, each with a cone half-angle of $35^{\circ}$.

  The main sources of background to the $\tp\tm$ selection are
        Bhabha events, dimuon events, multihadron events, and
        two-photon events.  This analysis uses the  standard OPAL
        $\tt$  selection \cite{tausel}, which was developed to identify
        $\tp\tm$ pairs and to 
        remove these backgrounds, with slight modifications
        to  further reduce Bhabha background in
        the $\tt$ sample.  

For each type of background remaining in the $\tt$ sample, a variable
        was chosen in which the signal and background can be
        visibly distinguished.
        The relative proportion of background
        was enhanced by loosening criteria which would normally
        remove that particular type of background from the $\tt$
        sample, and/or by applying further
        criteria to reduce the contribution from signal and to remove other
        types of background.  A comparison of the data and Monte Carlo
        distributions was then used to assess the modelling of the
        background.  In most cases, the Monte Carlo simulation was
        found to be consistent with the data.  When the data and Monte
        Carlo distributions did not agree, the Monte Carlo
        simulation  was adjusted to fit the data.  Uncertainties of 4\% to 20\% were
        assigned to the background estimates as a result of these
        comparisons.  

        The $\tt$ selection leaves a sample of 96,898 candidate $\tt$
        events, with a predicted fractional background of $(0.01055
        \pm 0.00052)$.  The backgrounds in the $\tt$ sample are
        summarised in Table \ref{ttausel}.  The errors shown in the
        table include both the statistical and systematic uncertainties.
The effect of a small bias in this event selection is discussed in Section \ref{systs}.
\begin{table} 
\begin{center}
\caption{\label{ttausel}
Fractional backgrounds in the $\tt$ sample.}
\begin{tabular} {lc} \hline
Background      & Contamination \\ \hline
$\reeee$        & $0.00305 \pm 0.00027$  \\
$\reemm$        & $0.00108 \pm 0.00022$ \\
$\reeqq$        & $0.00377 \pm 0.00015$ \\
$\reeeemm$      & $0.00108 \pm 0.00022$ \\ 
$\reeeeee$      & $0.00157 \pm 0.00028$  \\ \hline
Total           & $0.01055 \pm 0.00052$ \\ \hline
\end{tabular}
\end{center}
\end{table}

\section{\label{taumsel}The $\taum$ selection}
After the $\tp\tm$ selection, each of the 193,796 candidate $\tau$ jets is analysed
individually to see
whether it exhibits the characteristics of the required $\taum$ signature. A muon from a $\tau$ decay will result in a track in the central
tracking chamber, little energy in the electromagnetic and hadronic
calorimeters, and a track in the muon chambers.  The $\taum$ selection
is based  primarily on information from the central tracking chamber
and the muon chambers.  Calorimeter information is not used in the
main selection, but instead is used to
create an independent $\taum$ control sample that is used to evaluate
the systematic error in the Monte Carlo efficiency prediction.  
The selection criteria which are used to identify the $\taum$ candidates 
are described below.

The $\taum$ candidates are selected from jets with one to three tracks
in the tracking chamber, where the tracks are ordered according to
decreasing particle momentum.  The highest momentum track is
assumed to be the muon candidate.

\begin{figure*}
\mbox{\epsfig{file=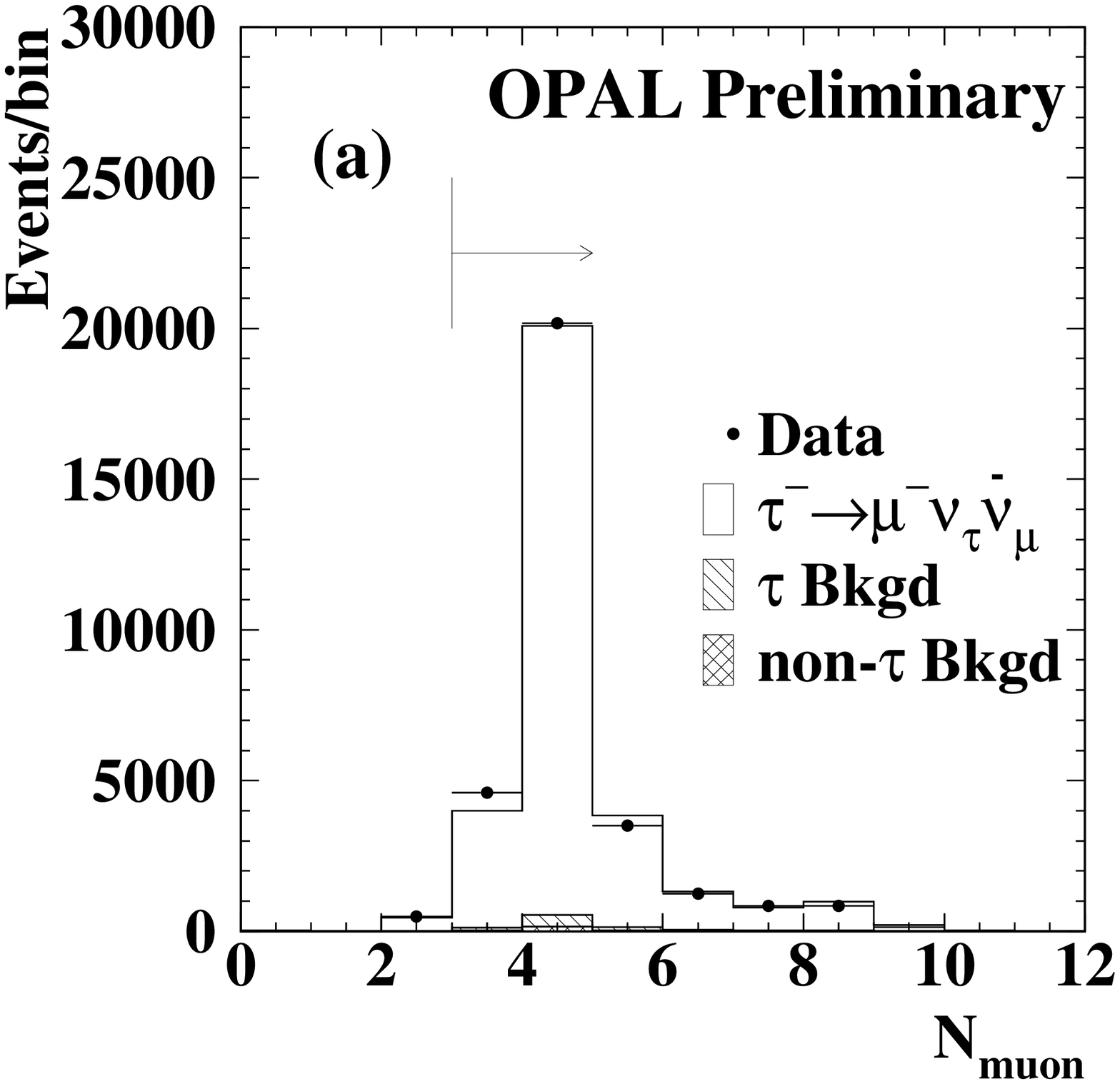,height=8cm}}
\mbox{\epsfig{file=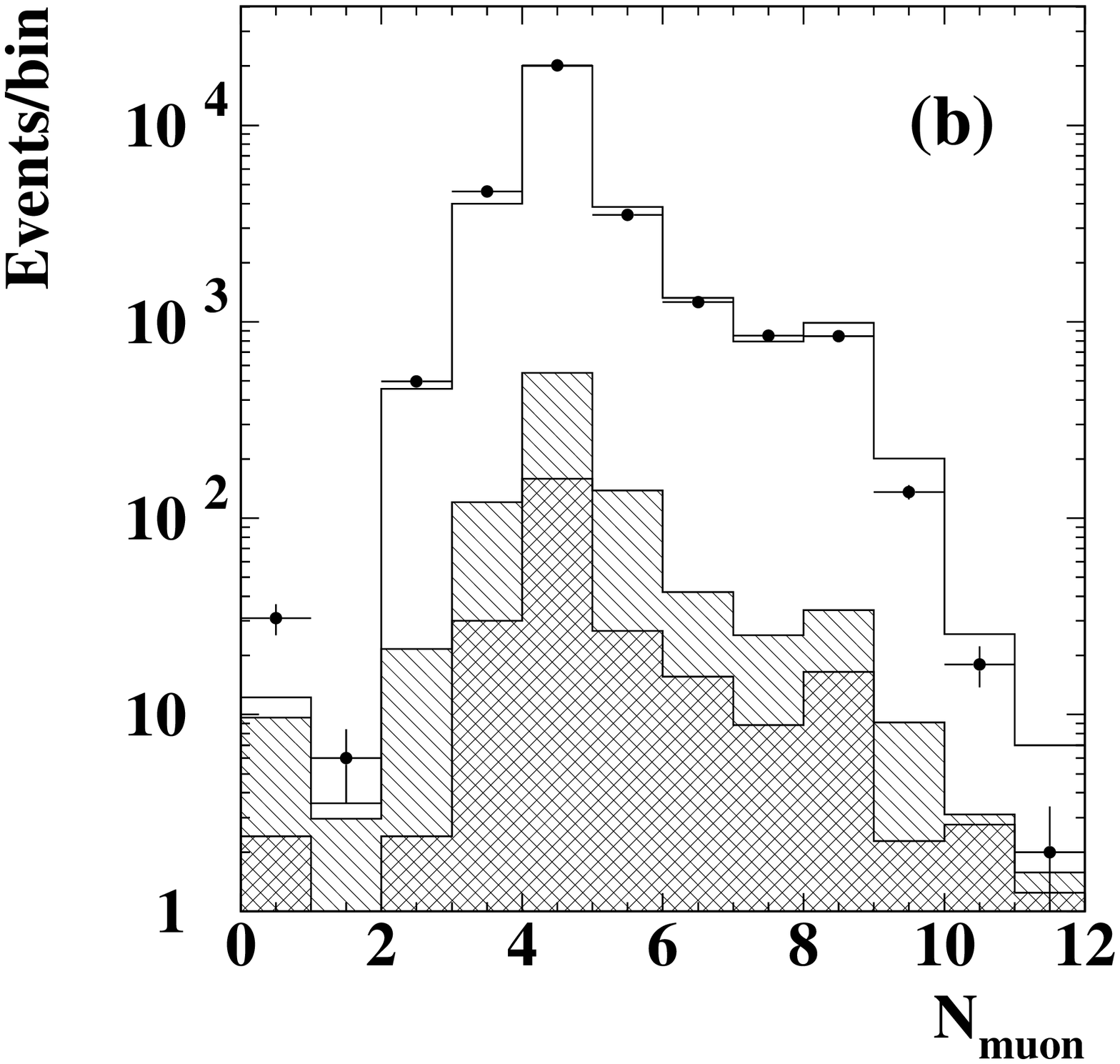,height=8cm}}
\mbox{\epsfig{file=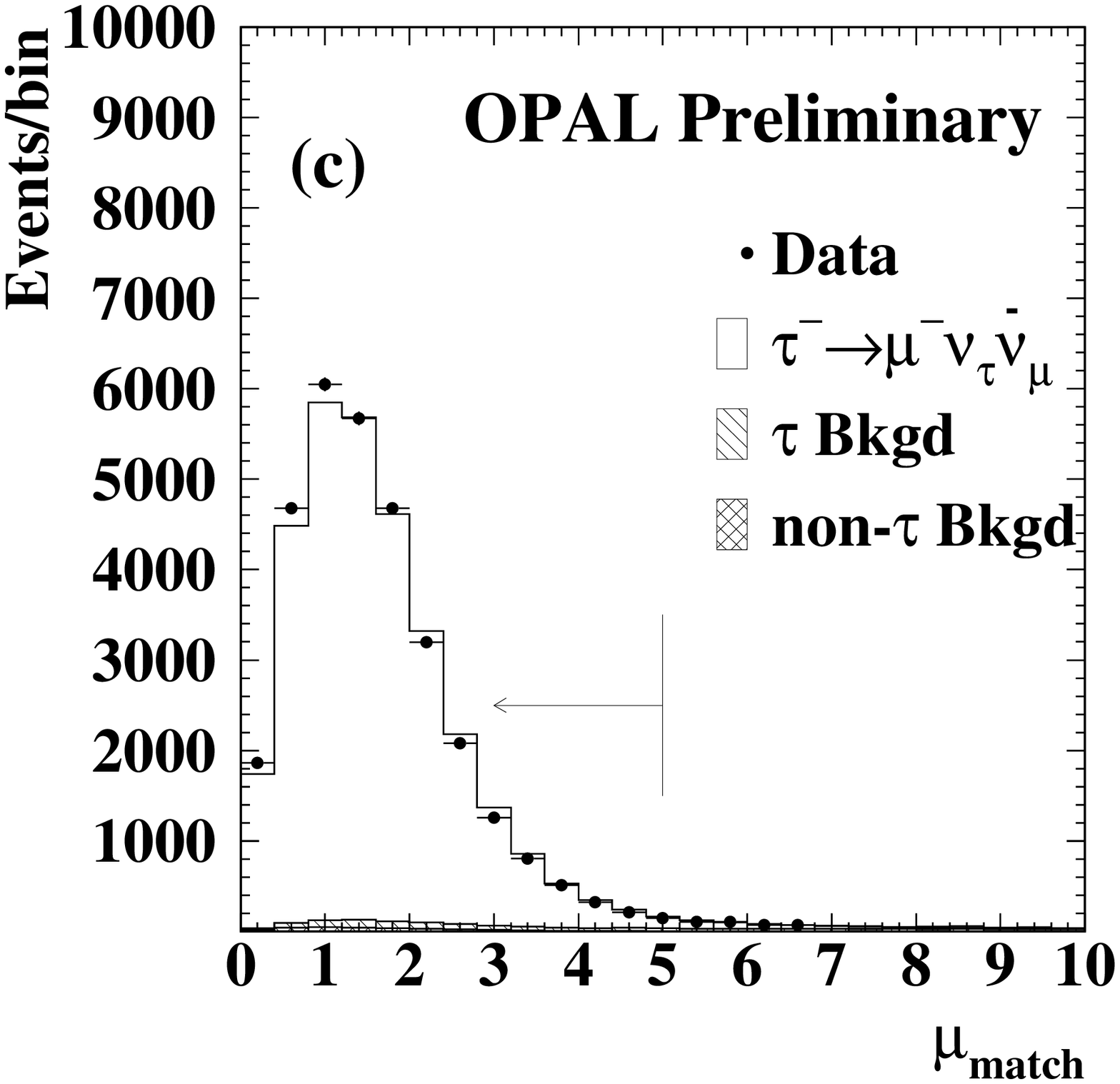,height=8cm}}
\mbox{\epsfig{file=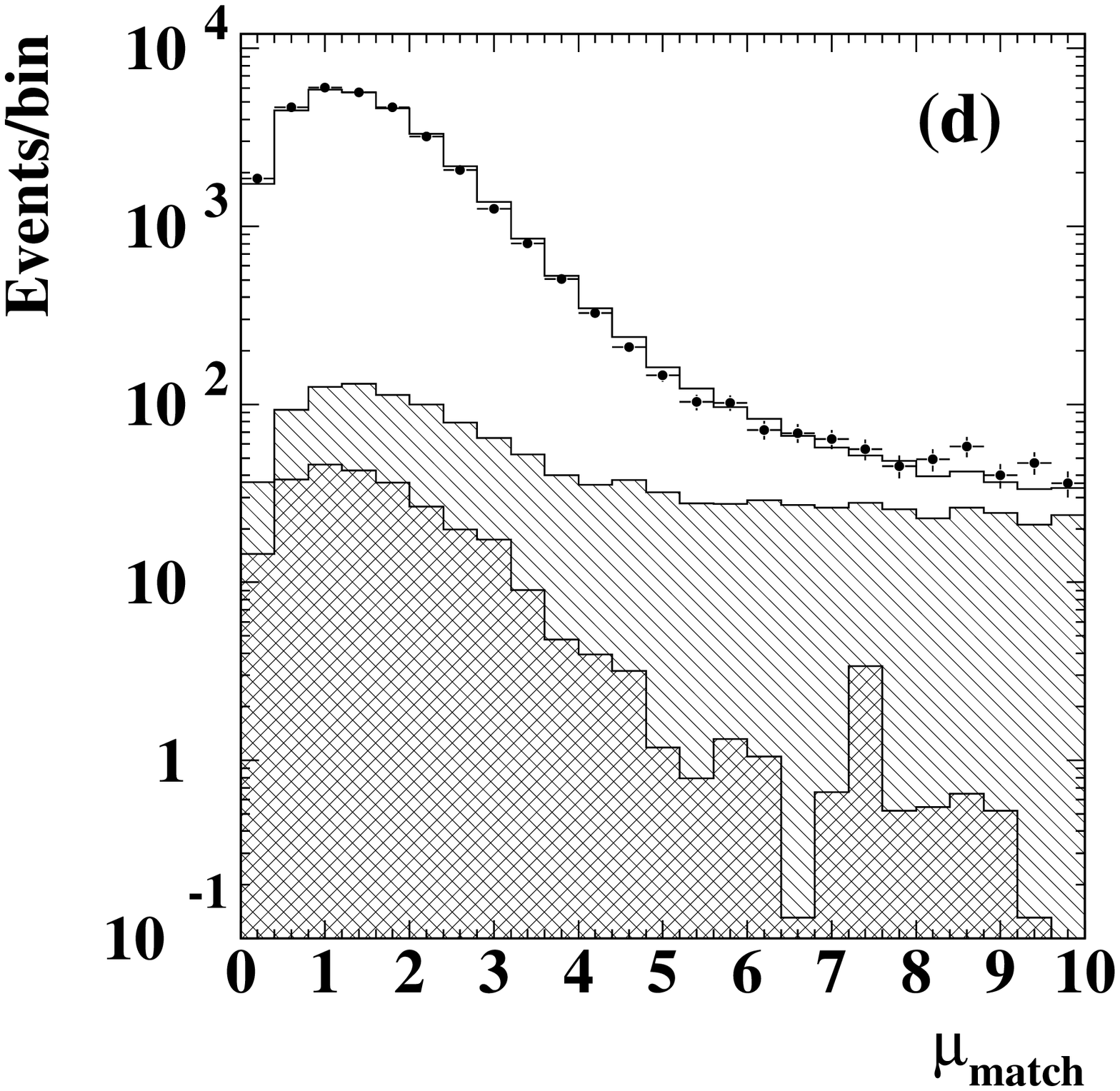,height=8cm}}
\caption{\label{cuts} ((a)
  and (b)) The
  number of muon layers, $\nmu$, activated by the passage of a charged
particle in the jet , and ((c) and (d)) the $\mumatch$
matching between a muon track reconstructed in the jet chamber
and one reconstructed in the muon chamber.  The jets in each
plot have passed all other selection criteria. The arrows indicate the
accepted regions.  The points are data, the clear histogram
is the Monte Carlo $\taum$ prediction, the singly-hatched histogram is
the Monte Carlo prediction for backgrounds from other $\tau$ decays,
and the cross-hatched histogram is the Monte Carlo prediction for
background from non-$\tau$ sources.}
\end{figure*}

Muons are identified as producing a signal in
at least three muon chamber layers, i.e. $\nmu > 2$, where $\nmu$ is
the number of muon chamber layers activated by a passing particle, as shown in Figure
\ref{cuts} (a) and (b).  The value of the $\nmu$ cut was chosen to
minimise the background while retaining signal.


  Tracks in the muon
chambers are reconstructed independently from those in the tracking
chamber.  The candidate muon track in the tracking chamber
is typically well-aligned with the corresponding track in the muon
chambers, whereas this is not the case for hadronic $\tau$ decays,
which are the main source of background in the sample.  The majority
of these background jets contain a pion which interacts in the hadronic
calorimeter, resulting in the production of secondary particles which
emerge from the calorimeter and generate signals in the muon chambers, a
process known as pion punchthrough.
Therefore, a
``muon matching'' variable, $\mumatch$, which compares the agreement between
the direction of a track reconstructed in the tracking chamber and
that of the track
reconstructed in the muon chambers, is used to differentiate the signal
$\taum$ decays from
hadronic $\tau$ decays\footnote{$\mumatch$ determines the difference in
  $\phi$ and in $\theta$ between a track reconstructed in the tracking
  chamber and one reconstructed in the muon chambers.  The differences
  are divided by an error estimate and added in quadrature to form a
  $\chisq$-like comparison of the directions.}.
It is required that $\mumatch$ have a value
of less than 5, (see Figure \ref{cuts} (c) and (d)).  The position of
the cut was chosen to minimise the background while retaining signal.

\begin{figure}
\mbox{\epsfig{file=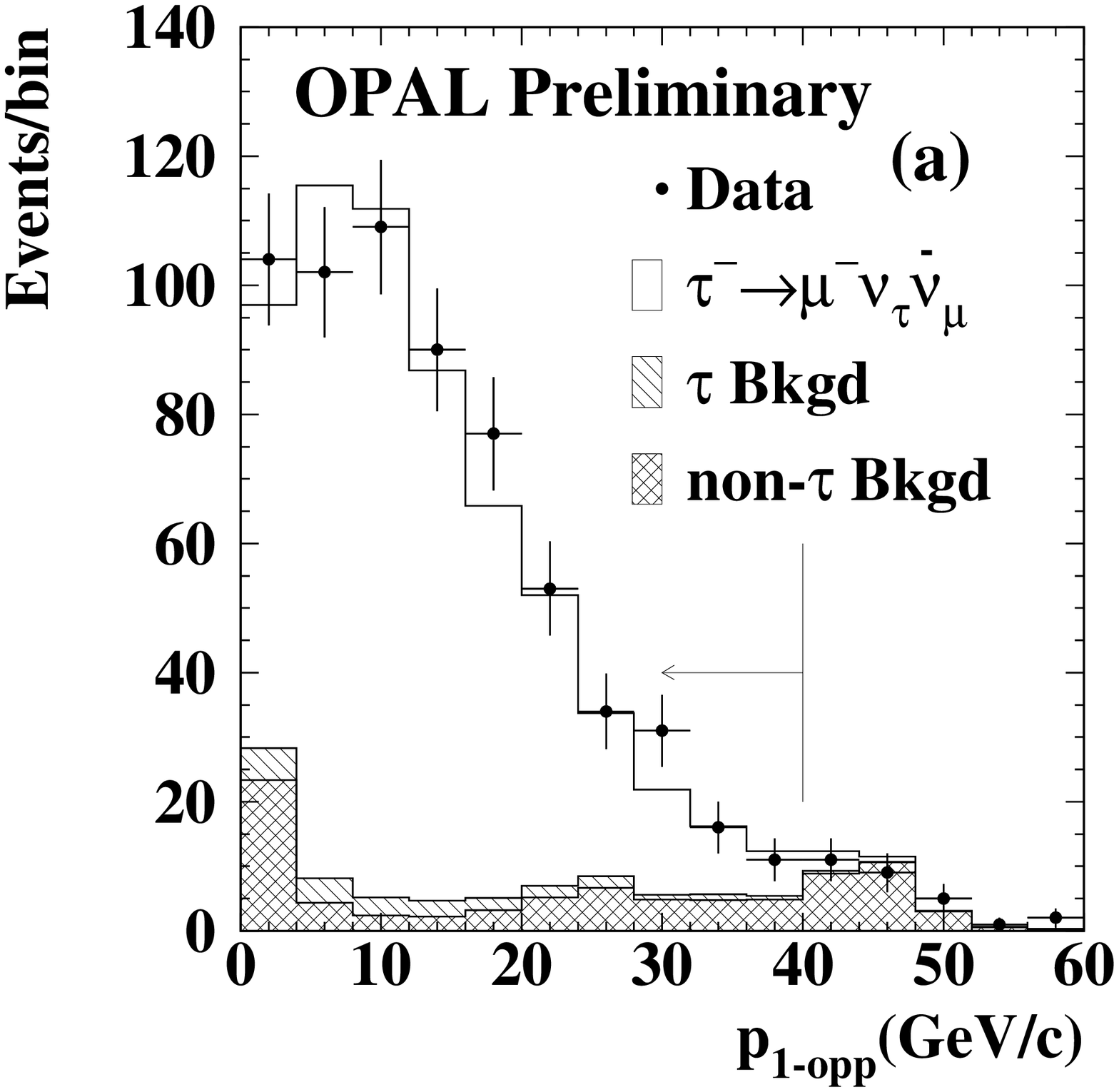,height=8cm}}
\mbox{\epsfig{file=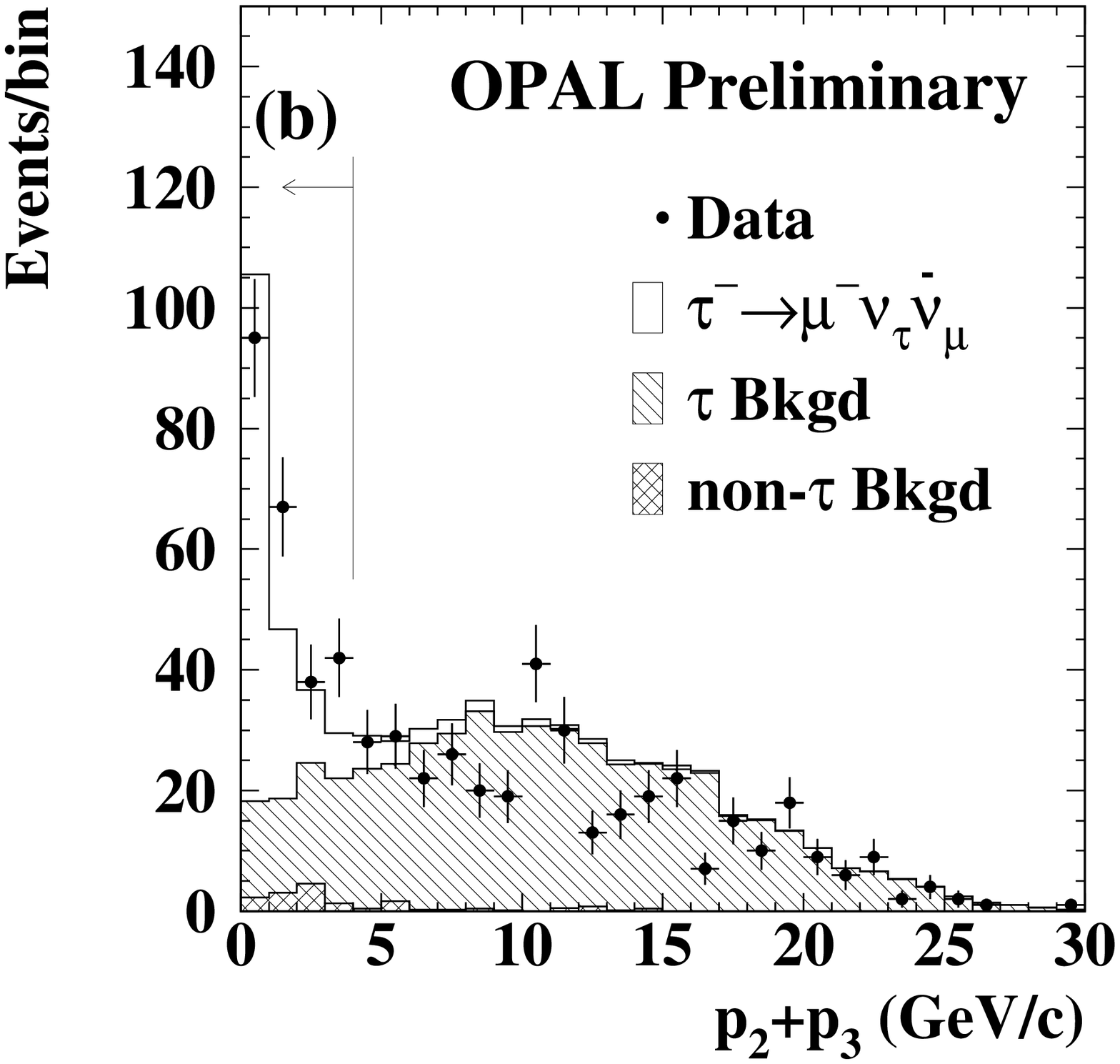,height=8cm}}
\caption[Momentum of 2nd and 3rd tracks.]{\label{ptks} (a) The momentum of
  the highest momentum particle in the opposite jet, $\popp$, where the
candidate muon has a momentum greater than 40 GeV/c, and (b) the combined momentum of the second and
third particles in the jet, for jets which have passed all
other selection criteria.  The arrows indicate the accepted regions.}  
\end{figure}

In order to reduce background from dimuon events, it is required that the
momentum of the highest momentum particle in at least one of the two
jets in the event, i.e. $\pa$ in the candidate jet and $\popp$ in the
opposite jet, must be less than 40 GeV/c.  (See Figure \ref{ptks} (a).)  

Although the $\taum$ candidates in general are expected to have one
 track, in approximately 2\% of these decays a radiated photon
 converts to an
 $\ee$ pair, resulting in one or two extra tracks in the
tracking chamber.  In order to retain these jets but eliminate
 background jets, it is required that the scalar sum of the 
momenta of the two lower-momentum particles, $\pb$ + $\pc$,  must be
 less than 4 GeV/c.  (See Figure \ref{ptks} (b).)  In cases where
 there is only one extra track, $\pc$ is taken to be zero.

\begin{figure}[h]
\mbox{\epsfig{file=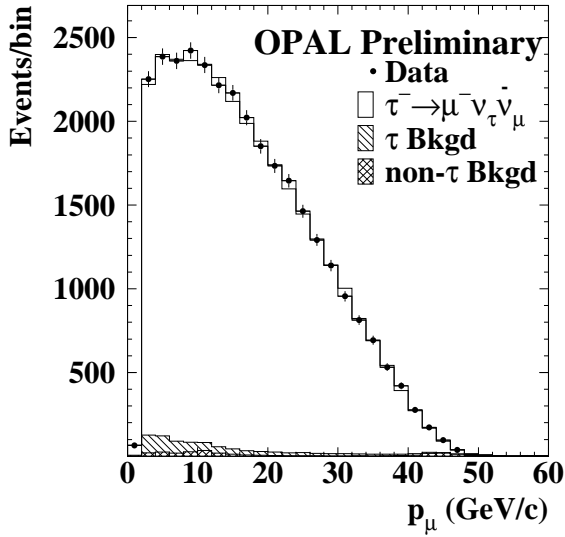,height=8cm}}
\caption{\label{ptk} The momentum of
  the candidate muon, for jets which have passed all of the selection criteria.}  
\end{figure}

The above criteria leave a sample of 31,395 candidate $\taum$ jets.
The backgrounds remaining in this sample are discussed in the next
section.  The quality of the data is illustrated in Figure \ref{ptk},
which shows the momentum of the candidate muon for jets which satisfy the $\taum$
selection criteria.

\section{\label{taumbk} Backgrounds in the $\taum$ sample}
\begin{figure*}
\mbox{\epsfig{file=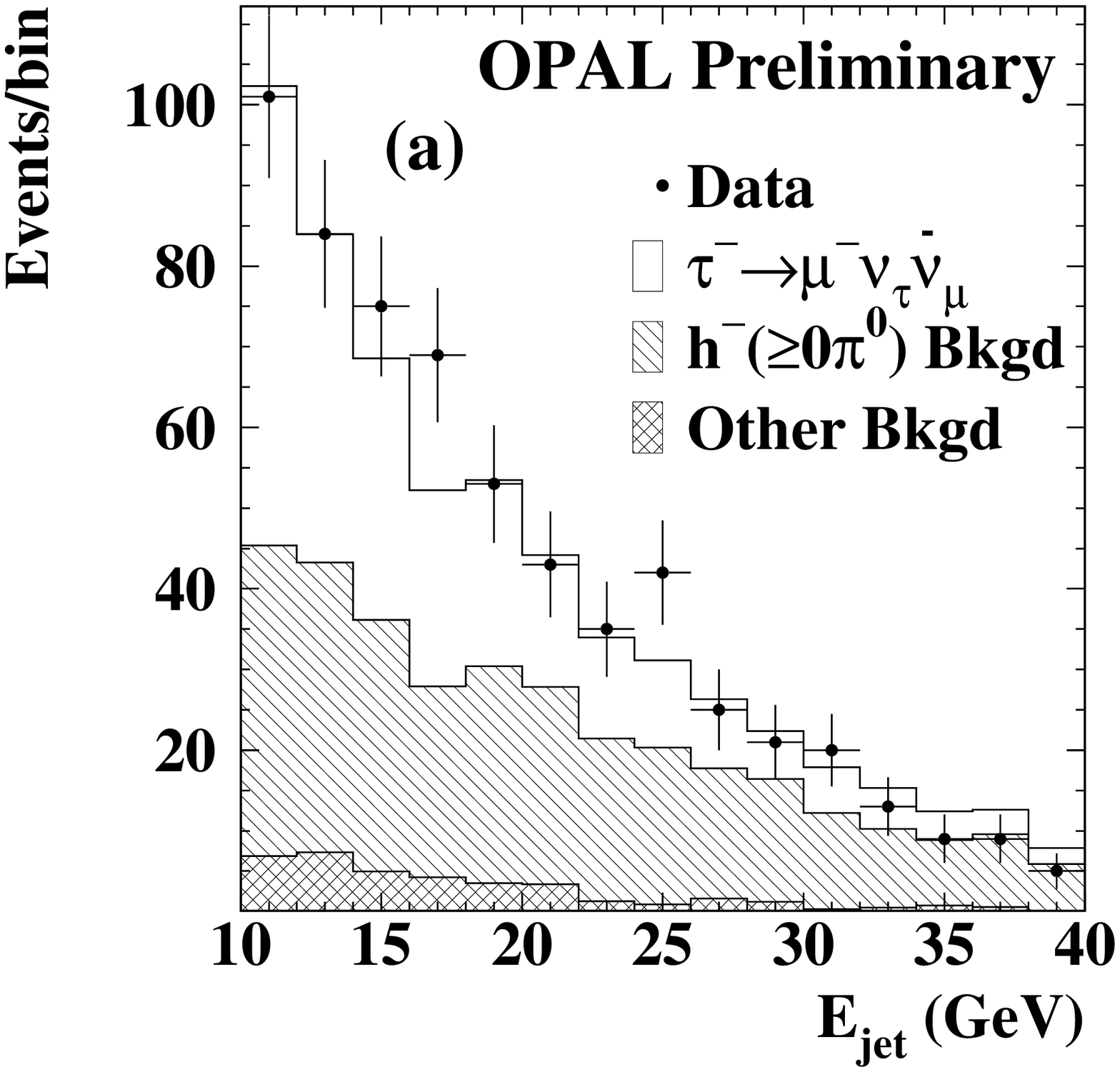,height=8cm}}
\mbox{\epsfig{file=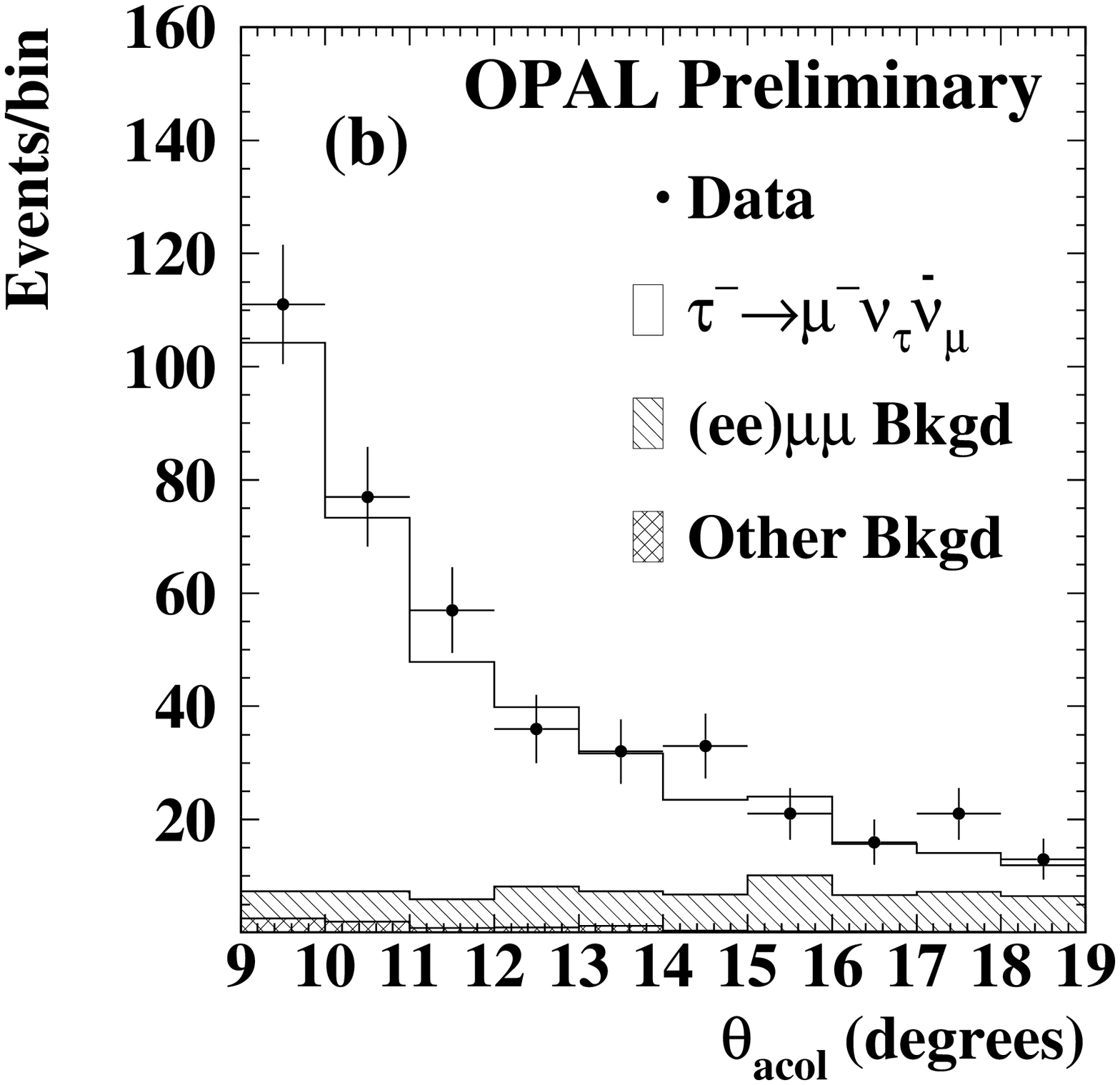,height=8cm}}
\mbox{\epsfig{file=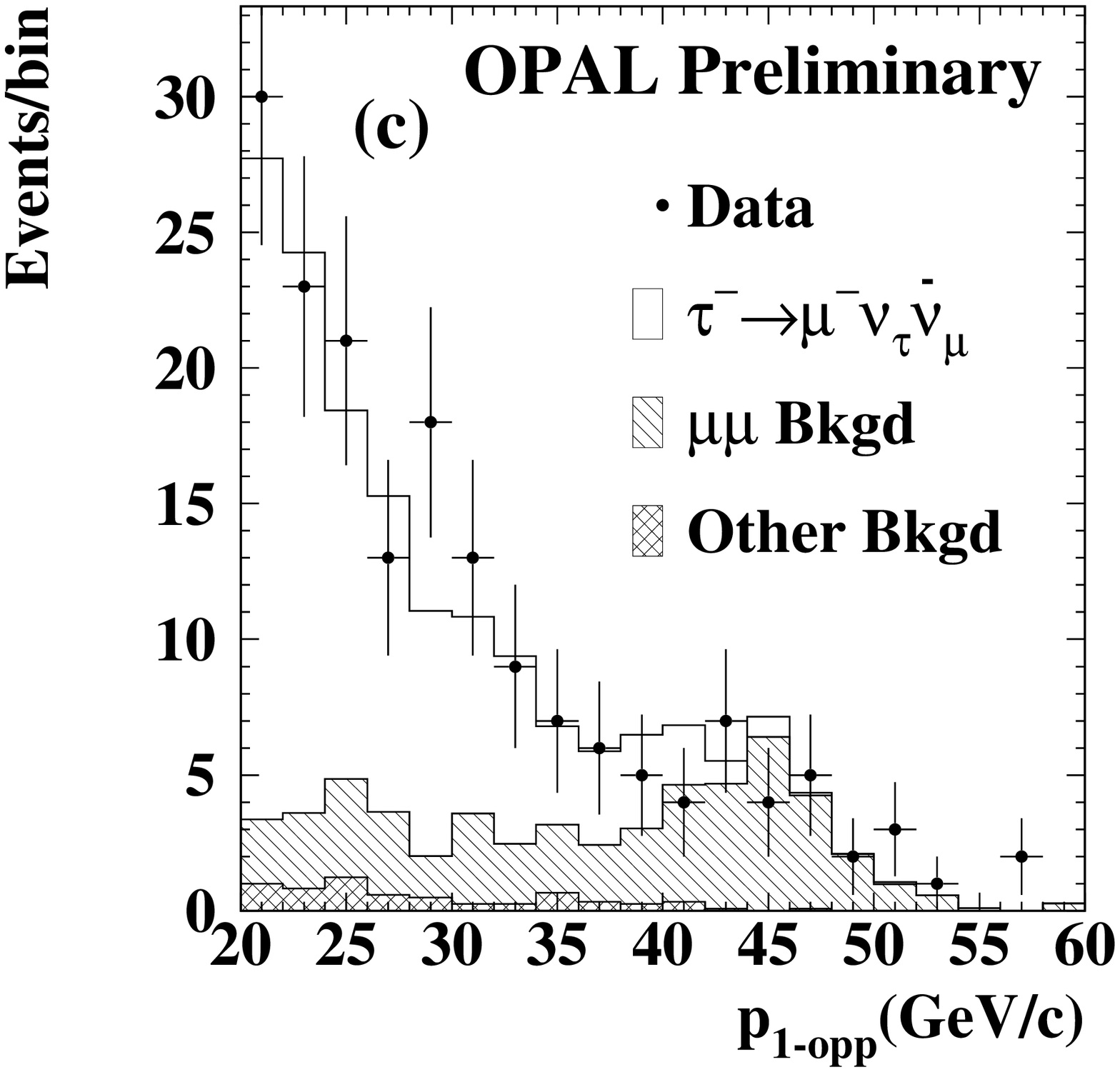,height=8cm}}
\mbox{\epsfig{file=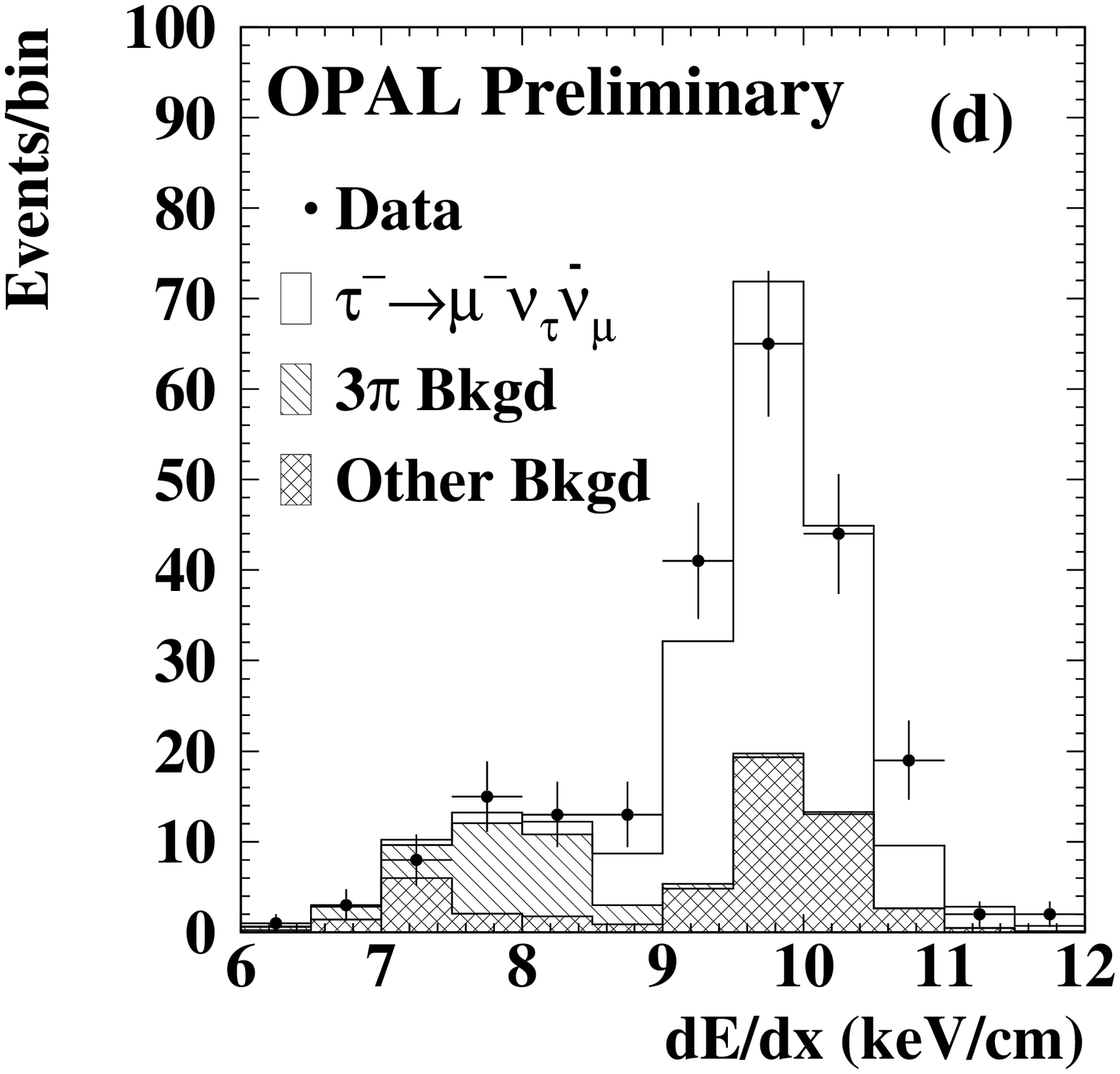,height=8cm}}
\caption{\label{bkdist} The
  distributions used to measure the background in the $\taum$ sample
  are shown. (a) $\ejet$ is the energy measured in the
  electromagnetic calorimeter, (b) $\acol$ is the acollinearity
  angle between the two $\tau$ jets, (c) $\popp$ is the momentum of the
  highest momentum particle in the opposite jet to the $\taum$
  candidate, (d) $\dedxt$ is the rate of energy loss of a particle
  traversing the tracking chamber.  The points are data, the clear histogram
is the Monte Carlo $\taum$ prediction, the singly-hatched histogram is
the Monte Carlo prediction for the type of background being evaluated
  using each distribution,
and the cross-hatched histogram is the Monte Carlo prediction for all
  other types of background.}
\end{figure*}

The background contamination in the signal $\taum$ sample stems from
other $\tau$ decay modes and from residual non-$\tau$ background in the
$\tt$ sample.  The procedure used to evaluate the background in the
$\taum$ sample is identical to the one used to evaluate the background
in the $\tt$ sample, which is outlined in Section \ref{tausel}, and
again involves studying the distributions of sensitive variables in
background-enhanced samples.  It is worth noting that, in all the plots in Figure \ref{bkdist}, the
relative proportion of background has been enhanced (as in Section
\ref{tausel}) in order that the background may be evaluated. The error
on each background includes the
Monte Carlo statistical error on the background fraction as well as
a systematic error associated with the background evaluation.

The main backgrounds from other $\tau$ decay modes can be
separated into $\tauhpmay$, and a small number of
$\tauhhh$ jets.  The $\tauhpmay$ decays can pass the $\taum$ selection
when the charged hadron punches through the calorimeters,
leaving a signal in the muon chambers.  The absence or presence of
$\pi^0$s has no impact on whether or not the jet is selected, since
there are over 60 radiation lengths of material in the detector in
front of the muon chambers.  The modelling of this background is
tested by studying $\taum$ jets with large deposits of energy in the
electromagnetic calorimeter. The distribution of jet energy,
$\ejet$,  deposited in the
electromagnetic calorimeter is shown in Figure \ref{bkdist} (a).  The
$\tauhpmay$ fractional background estimate is $0.0225 \pm
0.0016$, of which approximately 75\% includes at least one $\pi^0$.

The main backgrounds resulting from contamination in the $\tt$ sample are
$\reeeemm$ and $\reemm$ events.  The $\reeeemm$ contribution in the $\taum$ sample was
evaluated by fitting the Monte Carlo distribution of the acollinearity
angle, $\acol$,\footnote{Acollinearity is the supplement of the
        angle between the two jets.} to that
of the data, as shown in
Figure \ref{bkdist} (b). This
resulted in a fractional background estimate of $0.0052 \pm 0.0010$.  

The dimuon jets ($\reemm$) were enhanced in the $\taum$ sample by removing the
requirement that  $\popp < 40$ GeV/c or $\pa < 40$ GeV/c, and then requiring that
$\pa > 40$ GeV/c. The distribution of $\popp$ was then used to
evaluate the agreement between the data and the Monte Carlo simulation
for this background.  The resulting estimate of the dimuon fractional background in the
$\taum$ sample is $0.0029 \pm 0.0006$.  The corresponding 
distribution is shown in
Figure \ref{bkdist} (c).

The 3-track signal $\taum$ is due to photons which
convert in the tracking chamber to an $\ee$ pair, whereas the 3-track background
consists mainly of jets with three pions in the final state.
Electrons and pions have different rates of energy loss in the OPAL
tracking chamber, and hence the background can
be isolated from the signal 
by plotting the rate of energy loss as the particle traverses the tracking
chamber, $\dedxt$, of the second-highest-momentum particle in the
jet.  The Monte Carlo modelling was compared to the data as shown in
Figure \ref{bkdist} (d), yielding a fractional background measurement
of $0.0014 \pm 0.0003$.  

        The remaining background in the $\taum$ sample consists almost
        entirely of $\tau^- \rightarrow \pi^- \pi^0 \eta \nu_{\tau}$ jets,
        and contributes at a level of 0.04\% to the overall sample.  The total
        estimated fractional background in the $\taum$ sample after the 
        selection is $0.0324 \pm 0.0020$.  The main background
        contributions are summarised in Table
        \ref{tbackgrounds}.  The errors in the table include both the statistical and systematic uncertainties.

\begin{table} [h]
\begin{center}
\caption{\label{tbackgrounds}
The main sources of background in the candidate $\taum$ sample.}
\begin{tabular} {ll} \hline
Backgrounds     & Contamination  \\ \hline
$\tauhpmay$     & $0.0225 \pm 0.0016$ \\
$\reeeemm$      & $0.0052 \pm 0.0010$ \\
$\reemm$        & $0.0029 \pm 0.0006$ \\ 
$\tauhhh$       & $0.0014 \pm 0.0003$ \\
Other           & $0.0004 \pm 0.0001$ \\ \hline
Total           & $0.0324 \pm 0.0020$
\\ \hline
\end{tabular}
\end{center}
\end{table}

\section{\label{br}The branching ratio}

The $\taum$ branching ratio is given by
\begin{equation}
\label{ebr}
        \mbox{B} =  \frac{N_{(\tau \rightarrow \mu)}}{N_{\tau}} 
             \frac{(1-f_{\mathrm{bk}})}{(1-f_{\tau \mathrm{bk}})}
             \frac{1}{\epsilon_{(\tau \rightarrow \mu)}}
             \frac{1}{F_{\mathrm{b}}},
\end{equation}
where the first term,  $N_{(\tau \rightarrow
  \mu)}/N_{\tau}$, is extracted from the data and is the
number of $\taum$ candidates after the $\taum$ selection, divided by the number of $\tau$ candidates
selected by the $\tt$ selection.  The remaining terms in Equation
  \ref{ebr} are evaluated using Monte Carlo simulations.  These terms
  include the estimated fractional backgrounds in the $\taum$ and in the
  $\tt$ sample, $f_{\mathrm{bk}}$ and
  $f_{\tau \mathrm{bk}}$, respectively, which must be subtracted off the numerator and
  denominator in the first term of Equation \ref{ebr}.  The method by which these backgrounds are evaluated
  has been discussed in Sections \ref{tausel} and \ref{taumbk}.  The
  efficiency of selecting the $\taum$ jets out of the sample of $\tt$
  candidates is given by $\epsilon_{(\tau \rightarrow \mu)}$.  The Monte Carlo
  prediction of the efficiency is cross-checked using a control
  sample, and will be discussed in Section \ref{systs}.
  $F_{\mathrm{b}}$ is a bias factor 
which accounts for the fact that the $\tt$ selection does not select
all $\tau$ decay modes with the same efficiency.   The corresponding
values of these parameters for the $\taum$ selection are
shown in Table \ref{tbrvalues}.  Equation \ref{ebr} results in a
branching ratio value of 
\[
\mbox{\btaum} = 0.1734 \pm 0.0009 \pm 0.0005,
\] 
where the first error is
statistical and the second is systematic. 

\subsection{\label{systs} Systematic checks}

The statistical uncertainty in the branching ratio is taken to be the binomial error in the
uncorrected branching ratio, $N_{(\tau \rightarrow \mu)}/N_{\tau}$.  The systematic errors
include the contributions associated with the Monte Carlo modelling of
each of the main sources of background in the $\taum$ sample, the error
in the efficiency, the error in the
background in the $\tt$ sample, and the error in the bias
factor.  These
errors are listed in Table \ref{tbrvalues} and are discussed in more detail in the following paragraphs.

A second sample of $\taum$ data candidates was selected using
information from the tracking chamber plus the electromagnetic and hadronic calorimeters\footnote{Minimal tracking
  chamber information is used.  Specifically, this selection requires
  $1 \leq \ntks \leq 3$ and $\pb + \pc < 4$ GeV/c.}.  The candidates selected using this
{\it calorimeter} selection are highly correlated with those selected
for the main branching ratio analysis using the {\it tracking} selection, even though the two selection
procedures are largely independent.  Because of the high level of
correlation, the advantage of combining the two selection methods is
negligible; however, the calorimeter selection is very useful for
producing a
control sample of $\taum$ jets which can be used for systematic checks.  

A potentially important source of systematic error in the analysis is
the Monte Carlo modelling of the selection efficiency.  In order to
estimate the  error on the efficiency, both
data and Monte Carlo simulated jets are required to satisfy the
calorimeter selection criteria.  This produces two control samples of candidate $\taum$
jets,  one which is data, and one which is Monte Carlo simulation.
Each of these control samples is then passed through the
$\taum$  tracking selection.  The efficiency of the tracking selection is
then calculated for the data and for the Monte Carlo simulation by
taking the ratio of the number of jets in the control sample which pass
the selection, over the number of jets which were in the control
sample.  The difference between the efficiency determined using the
data and that determined using the Monte Carlo simulation was taken as the systematic error in the Monte Carlo
efficiency prediction.

Further checks of the Monte Carlo modelling were made by varying each of the selection
criteria.  In each case, the resulting changes in
the branching ratio were within
the systematic uncertainty that had already been assigned due to the background and
efficiency errors. 

The $\tau$ Monte Carlo simulations create events for the different $\tau$ decay modes, where the
proportions are determined by the branching ratios \cite{pdg}.  However, the $\tt$
selection does not select each mode of $\tau$ decay with equal
efficiency.  This can introduce a bias in the measured value of
B($\taum$).  The $\tt$ selection bias factor, $F_{\mathrm{b}}$, measures the
degree to which the $\tt$ selection favours or suppresses the decay
$\taum$ relative to other $\tau$ decay modes.  It is defined as the
ratio of the fraction of $\taum$ decays in a sample of $\tau$ decays
after the $\tt$ selection is applied, to the fraction of $\taum$
decays before the selection.  The uncertainty in $F_{\mathrm{b}}$ is
dominated by statistical error.

\begin{table} [h]
\begin{center}
\caption{\label{tbrvalues}
Branching ratio determination.}
\begin{tabular} {ll} \hline
Parameter       & Value     \\ \hline
$N_{(\tau \rightarrow \mu)}$    & 31,395 \\
$N_{\tau}$      & 193,796         \\
$f_{\mathrm{bk}}$        & $0.0324 \pm 0.0020$    \\
$f_{\tau \mathrm{bk}}$   & $0.0106 \pm 0.0005$  \\
$\epsilon_{(\tau \rightarrow \mu)}$     & $0.8836 \pm 0.0005$ \\ 
$F_{\mathrm{b}}$         & $1.034 \pm 0.002$     \\ \hline
B($\taum$)      & $0.1734 \pm 0.0010$ \\ \hline
\end{tabular}
\end{center}
\end{table}


\section{Discussion}

The value of B$(\taum)$ obtained in this analysis can be used in
conjunction with the previously measured OPAL value of B$(\taue)$ to
test various aspects of the Standard Model.  For example, the Standard Model assumption of lepton universality requires that the
coupling of the W particle is identical to all three generations of
leptons.  The leptonic $\tau$ decays have already provided some of the most
stringent tests of this hypothesis (see, for example, \cite{steve}).
With the improved precision of
B$(\taum)$ presented in this note, it is worth testing this
assumption again. In addition, the leptonic $\tau$ branching ratios can be used to
measure the Michel parameter $\eta$, which can be used 
to set a limit on the mass of the charged Higgs particle in the
Minimal Supersymmetric Standard Model.  These
topics are discussed below.

\subsection{Lepton universality}

The Standard Model predictions for the leptonic partial decay widths of the $\tau$ are given by \cite{tsai,marciano}
\begin{eqnarray}
\label{widlep2}
\lefteqn{\Gamma(\taul) =} \\
    &   \nonumber \left( \frac{\gt\gl}{8m^2_{\mathrm{W}}} \right)^2
                        \frac{m^5_{\tau}}{96\pi^3} \,\,                                               f\left(\frac{m^2_l}{m^2_{\tau}}\right)
                        (1+ \delta^{\tau}_{\mathrm{RC}}), 
\end{eqnarray}
where $l$ stands for $e$ or $\mu$, $m_l$ is the mass of the charged
daughter lepton in the $\tau$ decay, $m_{\tau}$ is the
mass of the $\tau$ particle, and $m_\mathrm{W}$ is the mass of the W
boson.   $g_{\tau}$ is the strength of the coupling between the $\tau$ particle and the W
propagator, and $g_l$ is the strength of the coupling of
the W to the daughter lepton $l$.
$f(m^2_l/m^2_{\tau})$ corrects for the masses of
the final state leptons, and $(1+ \delta^{\tau}_{\mathrm{RC}})$ takes
into account higher order corrections.

The Standard Model assumption of
lepton universality requires that the coupling constants $\ge$, $\gm$,
and $\gt$, are identical, thus the ratio $\gm/\ge$ is expected to be
unity.  This can be tested experimentally by taking the ratio of the
corresponding branching ratios, B$(\taum)$/B$(\taue)$.  The measured branching ratio is related to the
predicted partial decay width via the expression B$(\taul) =
\Gamma(\taul)/\Gamma_{\tau}$, where $\Gamma_{\tau}$ is the total
$\tau$ decay width, or the inverse of the $\tau$ lifetime.  Taking the
ratio of B($\taum$) and B($\taue$) yields
\begin{equation}
\label{egmge}
        \frac{\mbox{\btaum}}{\mbox{B}(\taue)}
                = C \, \frac{\gm^2}{\ge^2} 
\end{equation}
where $C = f(m^2_{\mu}/m^2_{\tau})/f(m^2_{\mathrm{e}}/m^2_{\tau}) =   0.9726$.  We use Equation \ref{egmge} to
compute the coupling constant ratio, which, with the value
of B($\taum$) from this work and
the OPAL measurement of B($\taue$) = $0.1781 \pm 0.0010$ \cite{steve},
yields
\begin{equation}
\label{egmgeresult}
        \frac{\gm}{\ge} = 1.0005 \pm 0.0043,
\end{equation}
in good agreement with expectation.

In addition, the expressions for the partial widths of the $\taum$ and $\mue$ decays can be rearranged to test lepton
universality between the first and third lepton generations, yielding
the expression
\begin{equation}
\label{egtge}
        \frac{\gt^2}{\ge^2} = \mbox{\btaum} \, \, 
        \frac{m_{\mu}^5}{m_{\tau}^5} \, \, 
        \frac{\tau_{\mu}}{\tau_{\tau}} \, \, 1.0278.
\end{equation}
  Using the OPAL value for
the $\tau$ lifetime, $\tau_{\tau} =
289.2 \pm 1.7 \pm 1.2$ fs \cite{ttau}, the BES collaboration value for
the $\tau$ mass, $1777.0 \pm 0.3$ MeV/$c^2$ \cite{bes}, and the Particle
Data Group \cite{pdg} values for the muon mass, $m_{\mu}$, and muon lifetime,
$\tau_{\mu}$, we obtain,
\[
        \frac{\gt}{\ge} = 1.0031 \pm 0.0047,
\]      
again in good agreement with the Standard
Model assumption of lepton universality.

\subsection{Michel parameter $\eta$ and the charged Higgs mass}

The most general form of the matrix element for $\tau$ leptonic decay involves all possible
combinations of scalar, vector, and tensor couplings to left- and
right-handed particles (see, for example, \cite{boyko}).  In the Standard Model, the  coupling terms 
take the following values:   $g^{\mathrm{V}}_{\mathrm{LL}} = 1$ and all other
$g^{\gamma}_{i,j} = 0$, where $\gamma$ = S, V, or T for scalar,
vector, or tensor couplings, and $i,j$ = L or R for the chirality of
the initial and final state charged leptons.  These coupling terms
represent the relative contribution of each particular type of coupling to the
overall coupling strength, $G_\mathrm{F}$.  

The shape of the $\tau$ leptonic decay spectrum is more
conveniently parameterized in terms of the four Michel parameters \cite{michel,rainer},
$\eta$, $\rho$, $\xi$, and $\delta$,  each of
which is a linear combination of all possible couplings
$g^{\gamma}_{i,j}$.   The integrated decay width is given
by 
\begin{equation}
\label{egamma}
\Gamma_l = \Gamma_l^{(\mathrm{SM})}(1 + 4 \eta \frac{m_l}{m_{\tau}}). 
\end{equation}

The Michel parameter $\eta$ is given by 
\begin{eqnarray}
\label{eeta}
\lefteqn{\eta = \frac{1}{2}
        Re\left\{g^\mathrm{V}_{\mathrm{LL}} \, 
        g^{\mathrm{S*}}_{\mathrm{RR}} + g^\mathrm{V}_{\mathrm{RR}} \,
        g^{\mathrm{S*}}_{\mathrm{LL}} \right.} \\
&\nonumber\left. + \, g^\mathrm{V}_{\mathrm{RL}}(g^{\mathrm{S*}}_{\mathrm{LR}} +
        6g^{\mathrm{T*}}_{\mathrm{LR}})
 +   g^\mathrm{V}_{\mathrm{LR}}(g^{\mathrm{S*}}_{\mathrm{RL}} +
        6g^{\mathrm{T*}}_{\mathrm{RL}})\right\},
\end{eqnarray}
and hence its Standard Model value is zero.  A
non-zero value of $\eta$ would affect the $\tau$ decay width via its
contribution to Equation \ref{egamma}. 
The term involving the ratio of masses in Equation \ref{egamma} acts as an effective
suppression factor in the case of $\taue$ decays; however, the
same is not true in $\taum$ decays.  It is possible then to solve for
$\eta$ by taking the ratio
$\Gamma_{\mu}(\eta)/\Gamma_\mathrm{e}(\eta)$, or equivalently by taking
the ratio of the measured branching ratios \cite{stahl}.  Using
Equations \ref{widlep2} and \ref{egamma}, we find  
\begin{equation}
\label{efindeta}
\frac{\mbox{\btaum}}{\mbox{B}(\taue)} = 0.9726 \left(1 + 4 \eta \frac{m_{\mu}}{m_{\tau}}\right)
\end{equation}
where $m_\mathrm{e}/m_{\tau}$ is taken to be zero and assuming lepton
universality ($\gm = \ge$).
The \btaum \, result presented here, together with 
the OPAL measurement of B$(\taue)$ \cite{steve} and Equation
\ref{efindeta}, then results in a value of $\eta = 0.004 \pm 0.036$.
This can be compared with a previous OPAL result of $\eta = 0.027 \pm
0.055$ \cite{rainer} which has been obtained by fitting the $\tau$ decay spectrum. 

In addition, if one assumes that the first term in the expression for $\eta$
is non-zero, then there must be a non-zero scalar coupling constant,
such that $\eta = \frac{1}{2}Re\{g^{\mathrm{S*}}_{\mathrm{RR}}\}$.  This coupling constant has been related to
the mass of a charged Higgs particle in the Minimal Supersymmetric Standard
Model via the
expression
        $g^\mathrm{S}_{\mathrm{RR}} = -m_l m_{\tau} (\tan\beta/m_{\mathrm{H}^{\pm}})^2,$ 
where $\tan\beta$ is the ratio of the vacuum expectation values of the two
Higgs fields.  $\eta$ can be approximately written as \cite{stahl}
\begin{equation}
\label{emasshiggs}
\eta = -\frac{m_{\tau}m_{\mu}}{2}
\left(\frac{\tan\beta}{m_\mathrm{H}^{\pm}}\right)^2.
\end{equation}
Thus, $\eta$ can be used to place constraints on the mass of the
charged Higgs, as has been done by Dova, Swain, and Taylor \cite{dova}.  A one-sided 95\% confidence limit using the
$\eta$ evaluated in this work gives a value of $\eta > -0.055$, and
a model-dependent limit on
the charged Higgs mass of $m_{\mathrm{H}^{\pm}} > 1.30 \tan\beta$.
This result is complementary to that from another recent OPAL analysis
\cite{btotau}, where a limit of $m_{\mathrm{H}^{\pm}} > 1.89
\tan\beta$ has been obtained using the process $b \rightarrow \tau^-
\bar{\nu}_{\tau} \mathrm{X}$. 

\section{Conclusions}

OPAL data collected at energies near the Z$^0$ peak have been used
to determine the  $\taum$ branching ratio, which is found to be
\[
\mbox{\btaum} = 0.1734 \pm 0.0009 \pm 0.0005,
\]
where the first error is statistical and the second is systematic.
This branching ratio, in conjunction with the OPAL
$\taue$ branching ratio measurement, has been used to verify lepton
universality at the level of 0.5\%.  In addition, these branching
ratios have been used to obtain a value for the Michel parameter
$\eta = 0.004 \pm 0.036$, which in turn has been used to place a
model-dependent limit on the mass of the charged Higgs boson, $m_{\mathrm{H}^{\pm}} > 1.30
\tan\beta$, in the Minimal Supersymmetric Standard Model.

\end{document}